\documentclass[%
reprint,
superscriptaddress,
onecolumn,
 amsmath,amssymb,
 aps,
]{revtex4-2}

\usepackage{graphicx}
\usepackage{dcolumn}
\usepackage{bm}
\usepackage{hyperref}
\usepackage{amssymb}   
\usepackage{amsmath}
\usepackage{epstopdf}
\usepackage[normalem]{ulem}
\usepackage{wrapfig, soul}
\usepackage{tabu,multirow}

\usepackage{xcolor}

\newcommand{\red}[1]{\textcolor{black}{#1}}

\usepackage[left]{lineno}
\usepackage{notoccite}
\newcommand{\beq}{ \begin{equation} }
\newcommand{\eeq}{ \end{equation} }
\newcommand{\beqs}{ \begin{eqnarray} }
\newcommand{\eeqs}{ \end{eqnarray} }

\begin{document}
\title{Vapor flux on bumpy surfaces: condensation and transpiration on leaves} 

\author{Sunghwan Jung} 
\email{sunnyjsh@cornell.edu}
 \affiliation{Biological and Environmental Engineering, Cornell University, NY 14853, USA}

\date{\today}

\begin{abstract}
Drop condensation and evaportation as a result of the gradient in vapor concentration are important in both engineering and natural systems. One of the interesting natural examples is transpiration on plant leaves. Most of water in the inner space of the leaves escapes through stomata, whose rate depends on the surface topography and a difference in vapor concentrations inside and just outside of the leaves. Previous research on the vapor flux on various surfaces has focused on numerically solving the vapor diffusion equation or using scaling arguments based on a simple solution with a flat surface. In this present work, we present and discuss simple analytical solutions on various 2D surface shapes (e.g., semicylinder, semi-ellipse, hair). The method of solving the diffusion equation is to use the complex potential theory, which provides analytical solutions for vapor concentration and flux. We find that a high mass flux of vapor is formed near the top of the microstructures while a low mass flux is developed near the stomata at the leaf surface. Such a low vapor flux near the stomata may affect transpiration in two ways. First, condensed droplets on the stomata will not grow due to a low mass flux of vapor, which will not inhibit the gas exchange through the stomatal opening. Second, the low mass flux from the atmosphere will facilitate the release of high concentrated vapor from the substomatal space. 
\end{abstract}


\maketitle 

\section{Introduction} \label{sec:intro} 
Drop condensation frequently occurs on plant leaves as both ambient vapor concentration and vapor flux are high at the leaf surface (e.g., early morning or after rain). This condensation process plays an important role in plant transpiration. The plant transpiration process (i.e., the exchange rate of water vapor through the stomata) depends on a difference in vapor pressures or concentrations between the sub-stomatal cavity and the air just outside the stomata \cite{Susann1959}. Especially, uneven leaf surfaces (e.g., trichomes or bumpy epidermal cells) could alter the vapor concentration and its flux in the air. Therefore, the transpiration rate can vary depending on the topography on the leaf surface and vapor concentration outside the stomata. Recent work inspired plant surface structures showed variations in vapor flux for drop condensation or ice nucleation over different surface structures \cite{Yao2018,Yao2020,Park2016}.

In engineering systems, drop condensation is a key factor in designing heat transfer devices \cite{Wang2012,Griffith1973,Aizenberg2013,Miljkovic2013}, e.g., thermal power generators, solar power plants, waste incineration, and water harvesting applications. Particularly, controlling the location and amount of condensed droplets is one of the main technical challenges. Inspired by leaf's hierarchical structures, extensive work has been done in terms of selective location of drop condensation \cite{Olceroglu2014,Enright2012,Mockenhaupt2008,Ghosh2014}, which further controls ice nucleation too \cite{Boreyko2016,Ahmadi2018}.   

Here is a brief summary of two key steps in the condensation process. 
First, the initial nucleation process happens at the nano scale and strongly depends on the wettability. Tiny droplets can be nucleated between nanowax tubules on a leaf. To formulate this process, the Gibbs free energy associated with condensation \cite{Fletcher1970}, is given as  
$\Delta G = - (RT/V_\mathrm{w}) \ln (p_\mathrm{vapor}/p_\mathrm{sat})$ where $V_\mathrm{w}$ is the molar volume of condensed water, which is balanced with the Gibbs energy to create a tiny droplet $(2\gamma  \cos \theta_\mathrm{Equil}/r$; $\gamma$ is the surface tension, $\theta_\mathrm{Equil}$ is the equilibrium contact angle, and $r$ is the pore or groove size of the surface). Therefore, the nucleation happens when  
\beq  
\frac{2 \gamma \cos \theta_\mathrm{Equil}}{r} < - \frac{RT}{V_\mathrm{w}} \ln \left( \frac{p_\mathrm{vapor}}{p_\mathrm{sat}} \right) = - \frac{RT}{V_\mathrm{w}} \ln \left[ \mathrm{Relative\,Humidity/100} \right] \, , \nonumber
\eeq
\red{where $R$ is the universal gas constant, $T$ is the temperature, $p_\mathrm{vapor}$ is the actual vapor pressure, and $p_\mathrm{sat}$ is the saturated vapor pressure }
This relation explains that the droplet nucleation easily occurs on a hydrophilic surface ($\cos \theta_\mathrm{Equil} >0$) at any groove sizes ($r$) in saturated air (${p_\mathrm{vapor}}/{p_\mathrm{sat}}>1$). Hierarchical double-layer roughness (i.e., nanowax and microbumps) is typical for leaf surfaces in nature (e.g., Lotus leaf \cite{Patankar:2004hj,Gao:2006ej}, Katsura tree leaf \cite{Kang2018}, and a recent review in \cite{Barthlott2017}). Especially, the nanowax might provide the groove lengthscale to initiate the droplet nucleation. 
Even in slightly unsaturated air, the drop can nucleate above a certain groove size. However, on a hydrophobic surface ($\cos \theta_\mathrm{Equil} <0$)), the droplet nucleation happens  only in saturated air and above the minimum pore size as $ \left( r>{2 \gamma \cos \theta_\mathrm{Equil}} \left[-\frac{RT}{V_\mathrm{w}} \ln \left( \frac{p_\mathrm{vapor}}{p_\mathrm{sat}} \right) \right]^{-1} \right) $ \cite{Jo2015}. In short, droplets likely nucleate on a hydrophilic surface even in unsaturated air, but not easily nucleate on a hydrophobic surface. 

\red{Second, the droplets will grow as more vapor diffuses onto after the droplet nucleation.} Hence, the vapor diffusion flux determines the growth rate of the nucleated droplets. Following Fick's law of diffusion, the diffusion flux can be expressed as $-D \nabla c$  where $D$ is the diffusivity, $\nabla$ is the spatial gradient, and $c$ is the vapor concentration. The diffusion flux is affected by the surface topography like leaf epidermal cells since the vapor concentration field is deformed due to a constant vapor pressure on the surface. Typically, this diffusion process is slow compared to the air convection, so it acts as a barrier for the exchange of vapor or other gases between the surface and the atmosphere. For example, a large portion of the total mass-transfer resistance from the atmosphere to the surface is attributed to the resistance by the boundary layer  \cite{Black1981,Choudhury1988,JamesCollatz1991}. Likewise, understanding the vapor concentration and diffusion flux in the presence of leaf’s microstructures will be an important task to understand the exchange of vapor between the leaf and the atmosphere. 

In this present study, we theoretically calculate the vapor concentration field and mass flux \red{on a non-hygroscopic surface at a constant temperature} to characterize the effect on vapor exchange through the stomata. First, we explain the analogies between flows past an object and vapor concentration fields around an object using the complex-potential theory. \red{Second}, we present solutions of the vapor concentration and diffusion flux on either flat, semicircular, semi-elliptical, or hair-like surfaces. Additionally, we calculate the vapor concentration and diffusion flux when a leaf stomatum opens in between two bumps. Finally, we discuss the biological benefits of leaf's microstructures in terms of plant transpiration.

\begin{figure}[t]
    \centering
    \includegraphics[width=.5\textwidth]{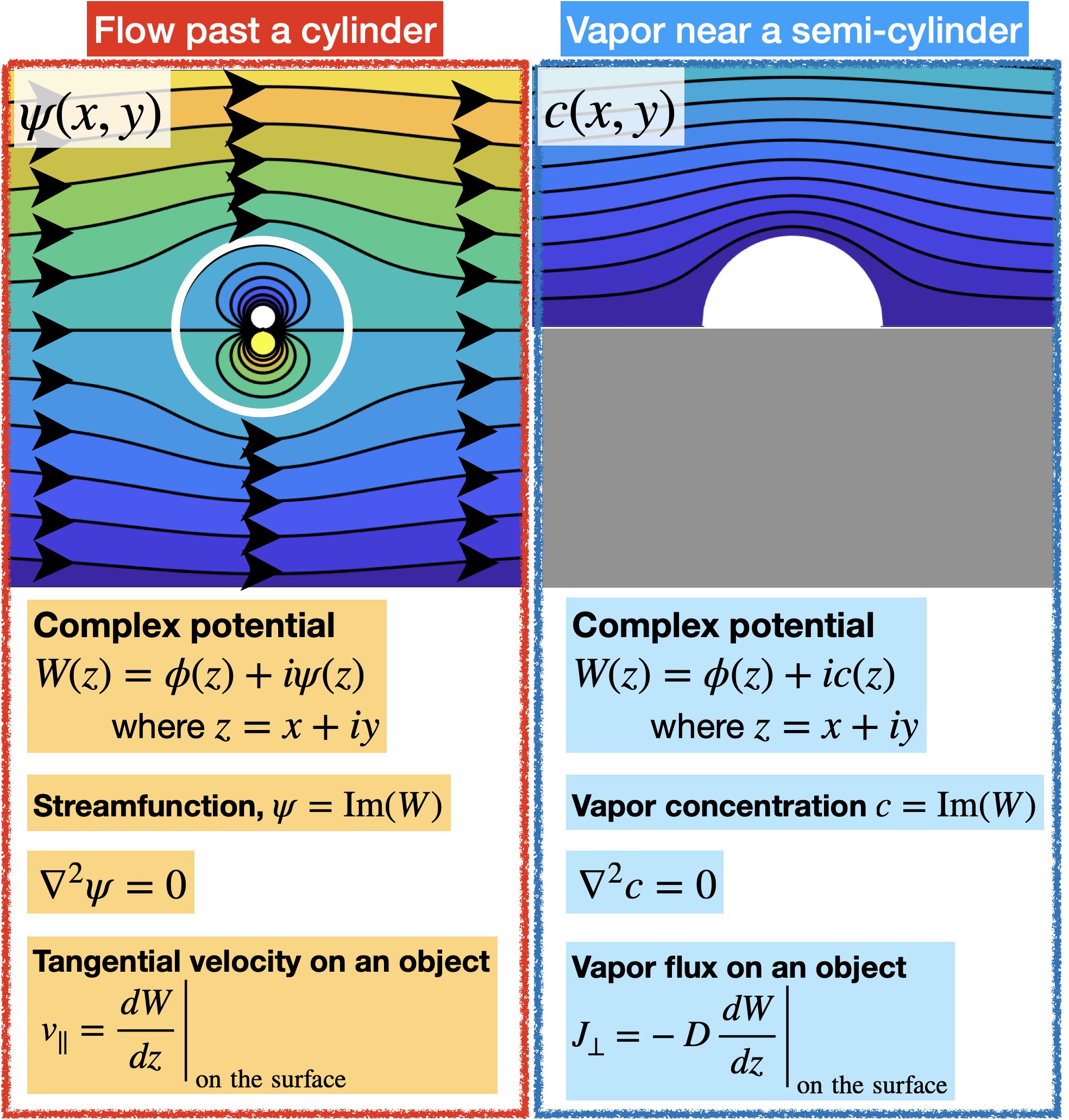}
    \caption{Similarity between a streamfunction and a vapor concentration around \red{a non-hygroscopic} object. }
    \label{fig:similarity}
\end{figure}

\section{Results}

\subsection{Similarity between a flow past an object and vapor concentration around an object}

One of the canonical examples in fluid dynamics is to describe an incompressible and irrotational flow past an object. Among many problem-solving methods, the complex potential theory is widely used in flow problems in aerodynamics, animal swimming \& diving \cite{Wu:2011ft,Chang:2016fg}, and hull slamming \cite{Abrate2013}. This complex potential method becomes more powerful along with conformal mapping to solve the flow solution around arbitrary shapes. There are many good textbook chapters for readers to study the basic concepts (Chap. 4 in ref. \cite{Acheson1990}, Chap. 4 in \cite{Currie2016}, Chap. 16 \& 17 in ref. \cite{Paterson1983}, Chap. 6 in ref. \cite{Kochin1964}). 
The complex potential is composed of a velocity potential as a real part and a streamfunction as an imaginary part. The irrotational condition leads to $\nabla^2 \psi(x,y) =0$ i.e., Laplace's equation. The flow velocity is given as ${\bf v} = (\nabla \times \psi) \cdot \hat{\bf b}$, where $\hat{\bf b}$ is the unit vector normal to the $x$-$y$ plane. 

We see a similar mathematical structure for the vapor concentration and its flux. The vapor concentration satisfies Laplace's equation ($ \nabla^2 c=0$) as a continuity condition $\nabla \cdot {\bf J}=0$ where the diffusion flux is ${\bf J} = -D \nabla c$. This diffusion flux corresponds to a mass flow from high to low vapor concentration regions. Likewise, we can calculate the concentration field and its flux mathematically using the complex potential method as we traditionally solve a fluid flow around an object (see Fig. \ref{fig:similarity}). Furthermore, we are able to estimate the preferred location and amount of condensation on various leaf surfaces. 


\subsection{Two canonical cases}
\red{\sout{Here,}} We consider two cases; a flat plate and a semicylinder on a plate. Let us consider a complex domain, i.e., the $z$-plane. Here, $z$ is defined as $x+iy$ where the real number corresponds to a $x$ coordinate and the imaginary number corresponds to a $y$ coordinate. \red{Both $x$ and $y$ coordinates are unitless values in this study. However, the coordinate dimensions can be interpreted relative to its bump size. }  

\subsubsection{Uniform flow past a plate or vapor on a plate} 
A basic example of a fluid flow is a uniform flow along a plate. We assume that the plate is aligned along the $x$ direction. \red{\sout{Here,}} All streamlines (contours of the streamfunction) are aligned along the $x$ direction thereby they are a function of $y$ coordinate only. Hence, when the flow speed is given as $U_0$, the complex potential, streamfunction in the bulk, and tangential velocity on the surface are given as 
\beq 
\frac{W(z)}{U_0} = z, ~ \frac{\psi}{U_0} = y, ~ \frac{ v_\parallel}{U_0} =   \frac{1}{U_0} \frac{\partial \psi }{ \partial y } = \frac{1}{U_0} \left. \frac{d W }{ dz } \right|_{y=0} =  1 \,.
\eeq
The same solution can be found for the case of the vapor on a flat surface. \red{Here,} the complex potential, concentration field in the bulk, and the diffusion flux on the surface are given as 
\beq
\frac{W^*(z)}{C_0/\xi_0} = z, ~ \frac{c}{C_0/\xi_0} = y, ~ \frac{J^*_\perp}{D C_0/\xi_0} = -\frac{1}{C_0/\xi_0} \frac{\partial c}{ \partial y } = -\frac{1}{C_0/\xi_0} \left. \frac{d W^* }{ dz } \right|_{y=0} = -1 \,,
\eeq
where $C_0$ \red{is the difference of the vapor concentration across a boundary layer, and $\xi_0$ is the boundary layer thickness. Hence, $C_0/\xi_0$ is the gradient of the vapor concentration over the boundary layer.} The diffusion flux on the surface, $J_\perp^*$, is a constant ($-D C_0/\xi_0$), which indicates that there is a uniform downward flux onto the surface.   

\subsubsection{Flow past a semicylinder or vapor around a semicylinder \label{sec:semicylinder}} 
The second canonical example is a flow around a cylinder, which is the same as the vapor concentration around a semicylinder as shown in Fig. \ref{fig:similarity}. The problem of a flow around a cylinder can be solved using a doublet (i.e., a dipole) added with a uniform flow. The complex potential, streamfunction in the bulk, and tangential velocity on the surface are given as 
\beqs 
\frac{W(z)}{U_0} &=& z \left(1 + \frac{a^2}{z^2} \right), ~\frac{\psi}{U_0} = r \sin\theta \left( 1 - \left( \frac{a}{r } \right)^2 \right), ~ \nonumber \\
\frac{ v_\parallel}{U_0} &=& \frac{1}{U_0} \left. \frac{dW}{ dz} \right|_{r=a} =  \sin \theta \left( 1 + \left( \frac{a}{r} \right)^2 \right) = 2 \sin\theta \,,
\eeqs
where $a$ is the radius of the cylinder. The first term in the complex potential describes a uniform flow along the $x$ direction and the second term represents a dipole flow. 
For the tangential velocity on a cylindrical surface, there are stagnation points ($v_\parallel=0$): front and back sides of the cylinder (i.e., $\theta = 0$ or $\pi$). The maximum velocity is achieved on the side ($\theta = \pi/2$).

This example is an analogy for the vapor concentration on a semicylinder on a flat surface. \red{ \sout{Here,}} The complex potential, concentration field in the bulk, and the diffusion flux on the surface are given as 
\beqs
&&\frac{W^*(z)}{C_0/\xi_0} = z \left(1 + \frac{a^2}{z^2} \right), ~\frac{c}{C_0/\xi_0}= r \sin\theta \left( 1 - \left( \frac{a}{r} \right)^2 \right), \nonumber \\ 
&& \frac{J^*_\perp}{D C_0/\xi_0} = -\frac{1}{C_0/\xi_0} \left. \frac{dW^*}{ dz} \right|_{r=a} = - \sin \theta \left( 1 + \left( \frac{a}{r} \right)^2 \right) = -2 \sin\theta \, . \label{eq:cyl}
\eeqs
This solution shows that the downward flux ($J_\perp^* < 0$) reaches the maximum value at the top of the bump ($\theta = \pi/2$) and becomes zero at the corners where the semicylinder meets the flat bottom surface ($\theta = 0$ or $\pi$). This trend can be observed from the contour lines as shown in the right panel of Fig. \ref{fig:similarity}. Densely (or sparsely) packed contours of vapor concentration represent a higher (or lower) diffusion flux. Therefore, more densely packed contours near the top of the semicylinder and less packed contours near the side corners indicate more vapor flux near the top and less flux near the side corners.

\subsection{Vapor flux with a single bump} 
We will consider the vapor concentration and diffusion flux on different shaped bumps beyond a simple semicylinder, e.g., prolate semi-ellipse and hair-like structures. From now on, we will use the normalized complex potential and vapor flux as $W \equiv W^*/(C_0/\xi_0)$ and $J_\perp \equiv J^*_\perp/(D C_0/\xi_0)$ for simplicity, respectively. \red{These normalized quantities can be understood as the values relative to the ones without any bump (i.e., a flat surface). }

\begin{table}[b]
    \centering
    \begin{tabular}{|c|c|c|c|}
    \hline
        Bump type & Normalized flux on the bump & Normalized flux outside the bump \\ \hline 
        Semi-cylinder  & $J_\perp = -2 \frac{y}{a} = -2\frac{\sqrt{a^2-x^2}}{a} , ~ |x|<a$ & $J_\perp = -\left( 1 - \frac{a^2}{x^2}\right), ~ |x|>a$ \\ \hline
        Prolate or oblate  &
        ${J_\perp} =  \frac{- 2(a^2\mp b^2) y}{ \sqrt{ (a\pm b^2/a)^4 x^2 + (a\mp b^2/a)^4 y^2 } }$ & ${J_\perp}= -1 + \left( 1 \pm \frac{a^2}{b^2} \right) \left( \frac{1}{2} - \frac{x/4}{\sqrt{ ( {x}/{2}  )^2 \pm b^2}} \right)$ \\ 
        semi-ellipse  &  
        when $|x| < a-b^2/a$ &  when $|x| > a-b^2/a$ \\ \hline
        Hair & $J_\perp = -\frac{y}{\sqrt{a^2 - y^2}}, ~ x = 0$ & $J_\perp = -\frac{x}{\sqrt{a^2 + x^2}}, |x|>0$ \\ \hline
    \end{tabular}
    \caption{Summary of normalized vapor flux around different single bumps}
    \label{tab:one}
\end{table}

\begin{figure*}
    \centering
    \includegraphics[width=.7\textwidth]{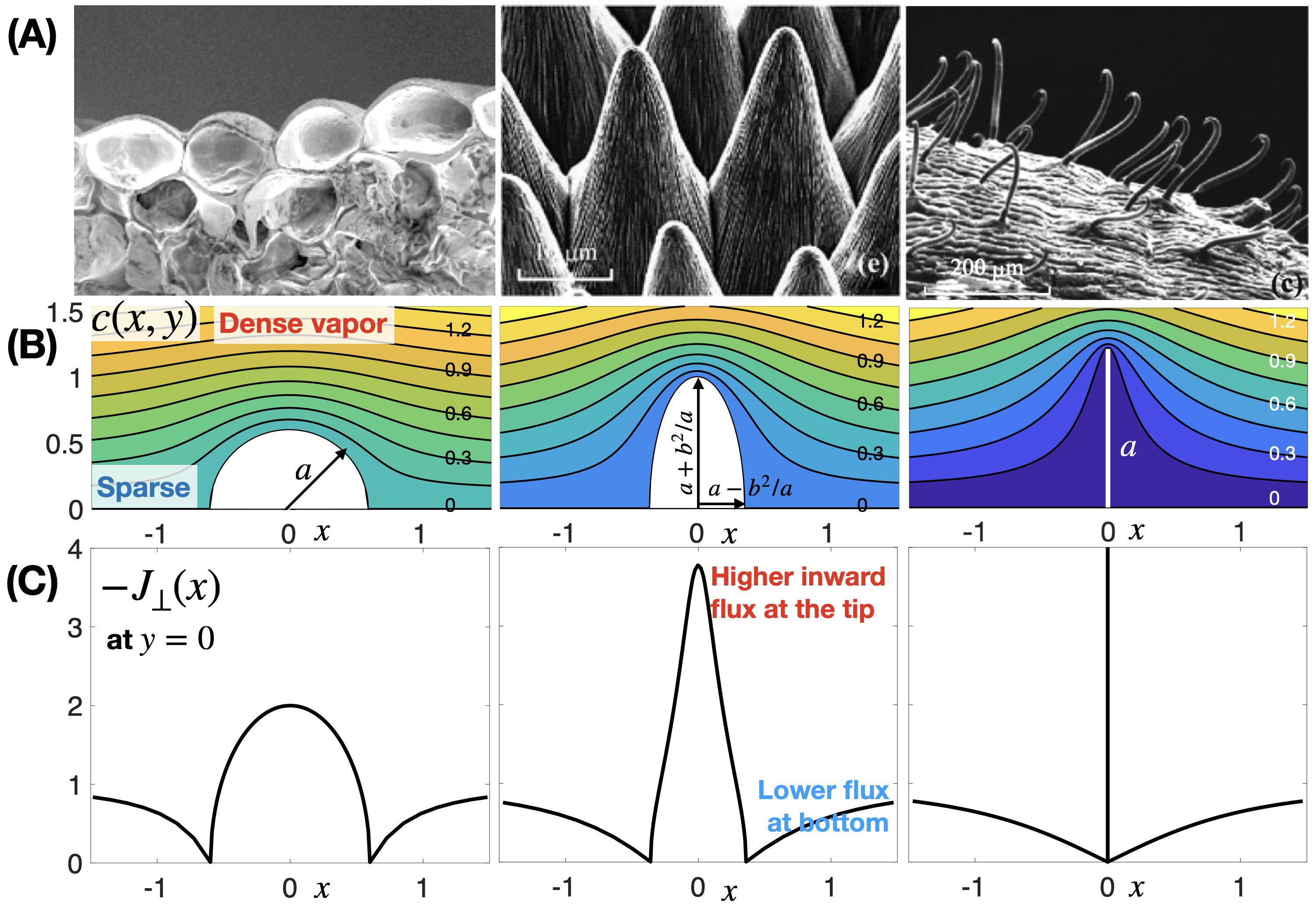}
    \caption{(A) Bump structures on plant surfaces (Images from left to right are \textit{Cercidiphyllum japonicum} \cite{Kang2018}; \textit{Viola tricolor} \cite{Barthlott2017}; \textit{Phaseolus vulgaris} \cite{Barthlott2017}; \red{the latter two images are reproduced from ref. \cite{Barthlott2017}}). (B) Vapor concentration fields near different bump structures: a semicylinder ($a=0.6$), a prolate semi-ellipse ($a=0.68$ \& $b=0.47$), and a hair ($a = 1.2$). (C) Three plots of downward diffusion flux normalized by the diffusion flux on a flat surface ($-J_\perp = D\nabla_\perp c$). As shown \red{ \sout{here}}, the highest diffusion flux (i.e., condensation growth rate) is observed on the top of the bump. }
    \label{fig:bump}
\end{figure*}

\subsubsection{Single semicylinder}
As described in the previous section, the vapor concentration and its derivative on a semicylinder with a flat surface are given as  
\beq
\mathrm{Im} [{W(z)}] = \mathrm{Im} \left[ z \left(1 + \frac{a^2}{z^2} \right) \right] \,, ~ \frac{dW}{dz} = \left( 1 - \frac{a^2}{z^2}\right) \, .
\eeq
The diffusion flux normal to the surface is simply a negative of the derivative of the complex potential. The reason is that the tangential component of the flux is always zero (like zero normal velocity in fluid flow) in the complex-potential method. Hence, the magnitude of the derivative equals to the flux normal to the surface ($J_\perp$). The diffusion flux on the semicylinder ($z = ae^{i\theta})$) is
\beq
J_\perp =  - \left.  \frac{dW}{dz} \right|_{r=a}  = -\sqrt{ (1-\cos 2\theta)^2 + \sin^2 2\theta} = - 2 \sin \theta = - 2 \frac{y}{a}\,. 
\eeq
We recover the same result as in Eq. (\ref{eq:cyl}) by replacing $y = a \sin \theta$. 

The diffusion flux outside the semicylinder ($z = x + i\cdot 0$) is 
\beq
J_\perp =  -\left. \frac{dW}{dz} \right|_{y=0 \mathrm{~and~} |x|\geq a} = -\left( 1 - \frac{a^2}{x^2}\right) \,. ~
\eeq
\red{\sout{As shown here,} This shows that} the diffusion flux is zero at the corners ($x=\pm a$) and slowly reaches to \red{$-1$} as $x$ gets far away from the corner. 

\subsubsection{Single semi-ellipse}
Conformal mapping is a powerful method to transform solutions of simple shapes (e.g., a plate and a semicylinder) into those in different complicated shapes. Using a transformation function, we can obtain solutions of vapor concentration and flux around a semi-ellipse. 
The transformation function from the $\zeta$ plane with a semicylinder to the $z$ plane with an ellipse is given as (similar calculations in p. 116\red{$-$}120 of ref. \cite{Currie2016})
\beq
z(\zeta) = \zeta \mp \frac{b^2}{\zeta} \mathrm{~~or~~} \zeta(z) = \frac{z}{2} + \sqrt{\left( \frac{z}{2} \right)^2 \pm b^2} \,,
\eeq
where $b$ is different from the radius of the semicylinder, $a$, in the pre-transformed $\zeta$ plane. The complex potential of a semi-ellipse is given as 
\beq
{W(z)} = \frac{z}{2} + \sqrt{\left( \frac{z}{2} \right)^2 \pm b^2} + \frac{a^2}{ \frac{z}{2} + \sqrt{\left( \frac{z}{2} \right)^2 \pm b^2} } = z - \left( 1 \pm \frac{a^2}{b^2} \right) \left( \frac{z}{2} - \sqrt{\left( \frac{z}{2} \right)^2 \pm b^2} \right) \,. \label{eq:cp_ellipse}
\eeq 
Here, a plus sign is for a prolate shape and a minus sign is for an oblate shape. The half length of the major axis becomes $a+b^2/a$ and the half length of the minor axis is $a-b^2/a$. The contour of the ellipse surface becomes 
$x = \left(a \mp {b^2}/{a} \right) \cos \theta, ~ y = \left(a \pm {b^2}/{a} \right) \sin \theta$ as a function of $\theta$, or $\left( {x}/({a \mp b^2 /a}) \right)^2 + \left( {y}/({a \pm b^2 /a}) \right)^2 = 1$ as a single equation.
The derivative of the complex potential is given as
\beq 
\frac{dW}{dz} = 1 - \left( 1 \pm \frac{a^2}{b^2} \right) \left( \frac{1}{2} - \frac{z/4}{\sqrt{\left( \frac{z}{2} \right)^2 \pm b^2} } \right)\,.
\eeq 

From the derivative, we can calculate the diffusion flux on the bump as 
\beq 
{J_\perp} =  -\frac{2(a^2\mp b^2) y}{ \left[ (a\pm b^2/a)^4 x^2 + (a\mp b^2/a)^4 y^2 \right]^{1/2}} \,. \label{eq:flux_ellipse}
\eeq 
Since the numerator linearly depends on $y$, we expect a higher downward flux at a higher $y$ (i.e., near the top of the bump), which has a similar trend of the solution of a semicylinder. Also, in the limit of $b\rightarrow 0$, the above solution converges to the solution on a semicylinder.  

The diffusion flux outside the bump along the flat surface $y = 0$ and $|x|>a \mp b^2/a$ is
\beq 
{J_\perp}= -1 + \left( 1 \pm \frac{a^2}{b^2} \right) \left( \frac{1}{2} - \frac{x/4}{\sqrt{ ( {x}/{2}  )^2 \pm b^2}} \right) \,.
\eeq 
In the limit of a position far from the bump (i.e., $|x| \gg 1$), the flux converges to minus one ($J_\perp \rightarrow -1$).  

\subsubsection{Single hair}
Using the conformal mapping, we can further solve the case with a hair-like structure.  
The transformation from the concentration around a semicylinder in the $\zeta$ plane to the concentration around a hair in the $z$ plane is (similar calculation in p. 136\red{$-$}139 of \cite{Currie2016}) 
\beq
z(\zeta) = \zeta - \frac{(a/2)^2}{\zeta}  \mathrm{~~or~~} \zeta(z) = \frac{z}{2} + \sqrt{\left( \frac{z}{2} \right)^2 + \left( \frac{a}{2} \right)^2}
\eeq
where $a$ is the height of the hair in the $z$ plane, which is the same as the radius of a circle in the $\zeta$ plane. 
Then, the complex potential is given as 
\beq
{W(z)} = \frac{z}{2} + \sqrt{\left( \frac{z}{2} \right)^2 - \left( \frac{a}{2} \right)^2 }  + \frac{a^2}{\frac{z}{2} + \sqrt{\left( \frac{z}{2} \right)^2 - \left( \frac{a}{2} \right)^2 } } = \sqrt{z^2 + a^2} \,.
\eeq 

Its derivative becomes  
\beq 
\frac{dW}{dz} = \frac{z }{\sqrt{z^2 + a^2}} \, .
\eeq 

The diffusion flux on the bump (i.e., $y\leq a$ and $x=0$; $z=0+iy$) is obtained from the derivative above.  
\beq 
{J_\perp} =  -\frac{y} {\sqrt{a^2 -y^2}}\,.
\eeq 
The flux increases close to the tip of the hair, but the solution will diverge at the tip $y\rightarrow a$.

The diffusion flux outside the hair along the surface (i.e., $y = 0$ \& $|x| > 0$; $z=x+i\cdot 0$) is
\beq 
{J_\perp} = -\frac{x}{\sqrt{a^2 + x^2}}\,.
\eeq 
It shows that the flux is zero at the corner and approaches to minus one in the far field. 

Figure \ref{fig:bump} summarizes our simulation results with single bumps of different shapes. Three cases are presented: a single semicylinder, a single prolate semi-ellipse, and a single hair. These structures are inspired by the microstructures found on real plant leaves (Circular epidermal bumps in \textit{Cercidiphyllum japonicum} \cite{Kang2018}; prolate bumps in \textit{Viola tricolor} \cite{Barthlott2017}; hair-like structures in \textit{Phaseolus vulgaris} \cite{Barthlott2017}). As shown in Fig. \ref{fig:bump}(B), vapor concentration contours are pushed up quite a bit with elongated prolate or hair-like structures. Therefore, its gradient (i.e., downward diffusion flux; $-J_\perp = \nabla c$) is high near the top of the bump and low at the lower side of the surface in Fig. \ref{fig:bump}(C).

\subsubsection{Total vapor flux on a single bump}
We will systematically study the trend of total diffusion flux for various shapes from oblate to prolate semi-ellipses. 
Since biomaterials are expensive to make, a fixed area would be a good criterion for systematic comparison here. The semi-ellipse has the half-length along the $x$ axis, $W \equiv a \mp b^2/a$, and the half-length (i.e., height) along the $y$ axis, $H \equiv a \pm b^2/a$. Then, the area of the semi-ellipse is a minor half-length times a major half-length as $(\pi/2)WH$. By choosing the reference area as a semicylinder with \red{a radius of $R_0$ as $(\pi/2)R_0^2$}, we can replace $W$ with \red{$R_0^2/H$}. Finally, the total flux is calculated by integrating $J_\perp$ over the bump surface. 

\red{Using the relation of $x^2/W^2 + y^2/H^2 =1$}, we rewrite the condensation flux of Eq. (\ref{eq:flux_ellipse}) on a semi-ellipse in terms of $W$ and $H$ as 
\beq 
{J_\perp} =  -\frac{W(W+H) y}{ \left[ H^4 x^2 + W^4 y^2 \right]^{1/2}} 
=  -\frac{(W+H) y}{ \left[ H^4 + (W^2-H^2) y^2 \right]^{1/2}}\,. \label{eq:Jperpy}
\eeq 
\red{The maximum flux always happens at $y=H$ as $J_\perp^{\mathrm{max}} = -(W+H)/W$.} 

\begin{figure*}
    \centering
    \includegraphics[width=.9\textwidth]{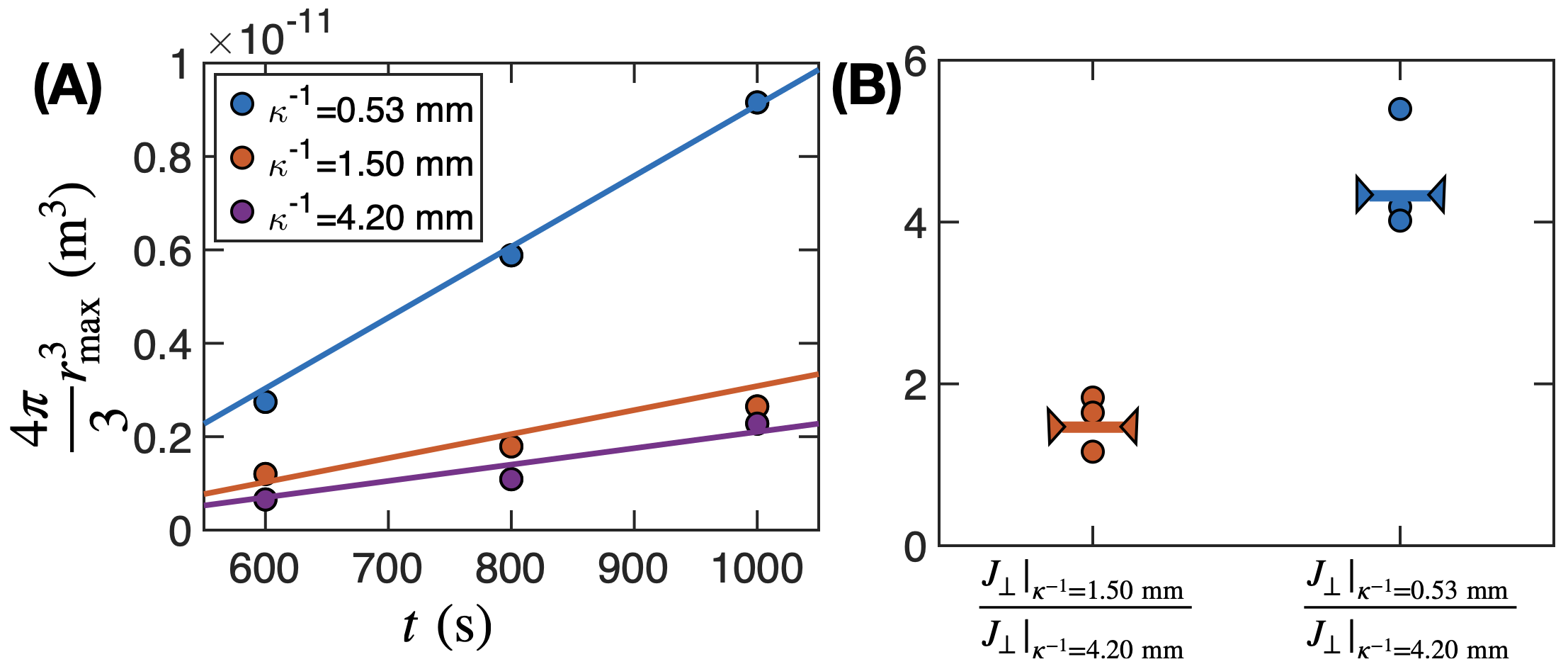}
    \caption{\red{ (A) Condensed droplet volume vs time. Circles are from previously published experiments in \cite{Park2016}. Different colors represent bumps with different curvatures. Solid lines are from our theoretical model as $\frac{4\pi}{3} r_\mathrm{max}^3 = 2\times10^{-15} \, J_\perp^\mathrm{max}  (t-400)$. Both prefactor of $2\times10^{-15}$ and time off-set of 300 are arbitrarily chosen. (B) Ratios of maximum diffusion fluxes. Circle symbols are from experiments \cite{Park2016}, and lines bounded by triangles are from our theory. The first group is a ratio of the diffusion flux with $\kappa^{-1}=1.50$ mm to one with $\kappa^{-1}=4.20$ mm.  The second group is a ratio of the diffusion flux with $\kappa^{-1}=0.53$ mm to one with $\kappa^{-1}=4.20$ mm. } }
    \label{fig:flux_max}
\end{figure*}

\red{Park et al. \cite{Park2016} performed condensation experiments on bumpy surfaces. In the experiments, they manufactured spherical caps with a fixed height ($H \simeq$ 0.8 mm) and three different radii of curvature ($\kappa^{-1} =$ 0.53, 1.50, and 4.20 mm). The bump width on the flat surface can be written in terms of the radius of curvature and the height as $W = H \sqrt{2\kappa^{-1}/H - 1}$. The corresponding maximum diffusion flux in 2D becomes $J_\perp^\mathrm{max} = -\left( 1+1/\sqrt{2\kappa^{-1}/H - 1} \right)$. 
Even though our theory is in 2D, we approximate the 3D maximum flux by taking the square of the 2D solution (similarly, the area is approximately $r$ in 2D and $r^2$ in 3D ). Therefore, the maximum flux on a 3D bump can be approximated as $J_\perp^{\mathrm{max}} \simeq -\left( 1+1/\sqrt{2\kappa^{-1}/H - 1} \right)^2$. Then, the measured volume of a condensed drop is expected to be proportional to the maximum diffusion flux multiplied by time. Figure \ref{fig:flux_max}(A) shows a plot of the condensed droplet volume versus the maximum flux mulitplied by time as $\frac{4\pi}{3} r_\mathrm{max}^3 \propto J_\perp^\mathrm{max} ~ t$. Experimental data points (circles) are well fitted with our theoretical models. Here, we arbitrarily choose the prefactor of $2\times10^{-15}$ and the time offset of 400 sec for all three cases. 
In addition, the ratio of the maximum diffusion fluxes is measured, which is not affected by these fitting parameters. Figure \ref{fig:flux_max}(B) shows that our predicted ratios of the maximum fluxes (lines bounded by two triangles) are in good agreement with experimental values (circles).  }

\begin{figure*}[t]
    \centering
    \includegraphics[width=1\textwidth]{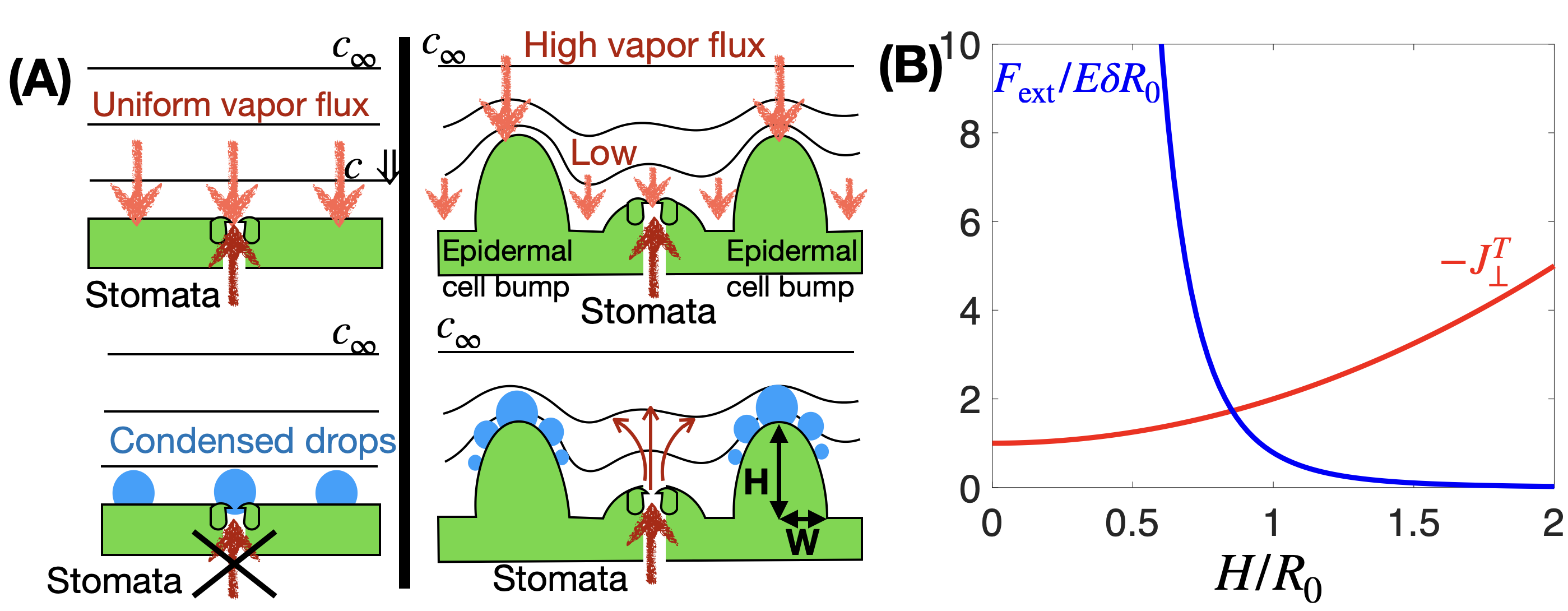}
    \caption{(A) Due to morphological features at the leaf surface, a low mass flux of vapor is developed near stomata while a high flux is formed near the top of epidermal cells. Due to the low vapor flux on the stomata will reduce the chance of clogging or reducing the stomatal opening. \red{Condensed drops on the lower panel are not part of initial conditions but rather a consequence of vapor flux.} (B) Normalized total flux and normalized bending force as a function of the normalized bump height. }
    \label{fig:flux_optimization}
\end{figure*}

If we integrate Eq. (\ref{eq:Jperpy}) from the bottom to the top of the bump ($0 \leq y\leq H$), then the total flux per unit area becomes
\beq 
{\cal J}_\perp^T =\frac{1}{W}\int_0^H J_\perp \sqrt{1 + \left( \frac{dx}{dy} \right)^2} dy =  \frac{1}{W}\int_0^H J_\perp \frac{ \sqrt{H^4 x^2 + W^4 y^2} }{H^2 x} dy = -\frac{(W+H)}{W} \,. 
\eeq 
This shows that the total flux increases as the bump has a more prolate shape ($H \uparrow$ and $W \downarrow$). 
A higher total flux by the bump is preferable for plants to avoid or inhibit condensation on the stomata that are usually located in between bumps or on the bottom of the leaf surface, not on the top of the bumps (see Figure \ref{fig:flux_optimization}(A)).

Even though a higher total flux is developed with a more prolate shape closer to a hair-like structure, a leaf may not maintain all prolate structures on the leaf surface. One of the major disadvantages of prolate or hair-like structures is structural instability or failure. If the structure is too elongated, then the structure can be easily bent or torn.  
For a semi-ellipse, the 2D bending second moment of the area, $I$, is proportional to $I = (\pi/4) W^3H$. Then, the resisting moment $F_\mathrm{ext} H$ is proportional to the bending rigidity $EI$, i.e., Young's modulus times the second moment of the area and the curvature, $\kappa$. The external force is proportional to $F_\mathrm{ext} = EI \kappa /H \approx (\pi/4) EI \delta /H^3 = (\pi/4) E \delta R_0^6 H^{-5}$ where $\delta$ is the horizontal deflection distance. 
\red{ \sout{As shown here,} This shows that} a narrower and higher bump ($H \uparrow$ and $W \downarrow$) can bend more and become vulnerable against external force, thereby destabilizing its structure on the leaf surface.  

Figure \ref{fig:flux_optimization}(B) shows the normalized total flux ($-{\cal J}_\perp^T$) and normalized bending force ($F_\mathrm{ext}/ E \delta$) as a function of its height, $H$. As the bump height increases, the total flux onto the bump becomes larger, but the bending force gets smaller. Presumably, plant leaf surfaces are evolved to optimize both higher mass flux and stable structures, thereby having a suitable height of bumpy microstructures. 

\subsection{Vapor flux with two bumps}

We consider the vapor concentration and its flux with two bumps on a flat surface. For potential flow calculations, the solution of a uniform flow through an array of cylinders can be obtained using Schwartz mapping \cite{Richmond1924}. More recently, D. Crowdy's group published a series of papers on this type of potential flow using conformal mapping or using Fourier transformation \cite{Crowdy2015,Crowdy2015a}. However, analytical expressions do not exist since there is no explicit (closed-form) expression of the transformation function and/or its inverse function. Instead, we propose a simple way to get an analytical solution of the vapor concentration and flux over two bumps. Proposed solutions can be obtained by superpositioning two complex potentials with a uniform flow potential. It is worth noting that our solution \red{{here}} is not an exact solution of two bumps, but instead approximates the vapor concentration and flux solutions around two semi-elliptical or semicircular bumps.

\begin{table}[b]
    \centering
    \begin{tabular}{|c|c|}
    \hline
        Bump type &  Normalized flux on bumps  \\ \hline
        Two semicylinders  & $ J_\perp = 1 - \left( \frac{(a+2L)^2 a^2}{(a+2L)^2+a^2} \right) \frac{2(z^2+L^2)}{(z^2-L^2)^2} $ \\ \hline
        {Two semi-ellipses}  &
        $ J_\perp = 1 - {\cal A}_1 \left( 1 \pm \frac{a^2}{b^2} \right)   \left(1 - \frac{\frac{(z-L)}{2^2}}{\sqrt{\frac{(z-L)^2}{2^2} \pm b^2}}  - \frac{\frac{(z+L)}{2^2}}{\sqrt{\frac{(z+L)^2}{2^2} \pm b^2}} \right)$  \\ 
        {} & {where ${\cal A}_1   =  \left[ {\left( 1 \pm \frac{a^2}{b^2} \right) \left(1 - \frac{\frac{a \mp b^2/a}{2^2}}{\sqrt{\frac{(a \mp b^2/a)^2}{2^2} \pm b^2}}  - \frac{\frac{a \mp b^2/a+2L}{2^2}}{\sqrt{\frac{(a \mp b^2/a+2L)^2}{2^2} \pm b^2}} \right)} \right]^{-1}$}
\\ \hline    \end{tabular}
    \caption{Summary of normalized vapor field and flux with two bumps}
    \label{tab:two}
\end{table}

\begin{figure*}
    \centering
    \includegraphics[width=.8\textwidth]{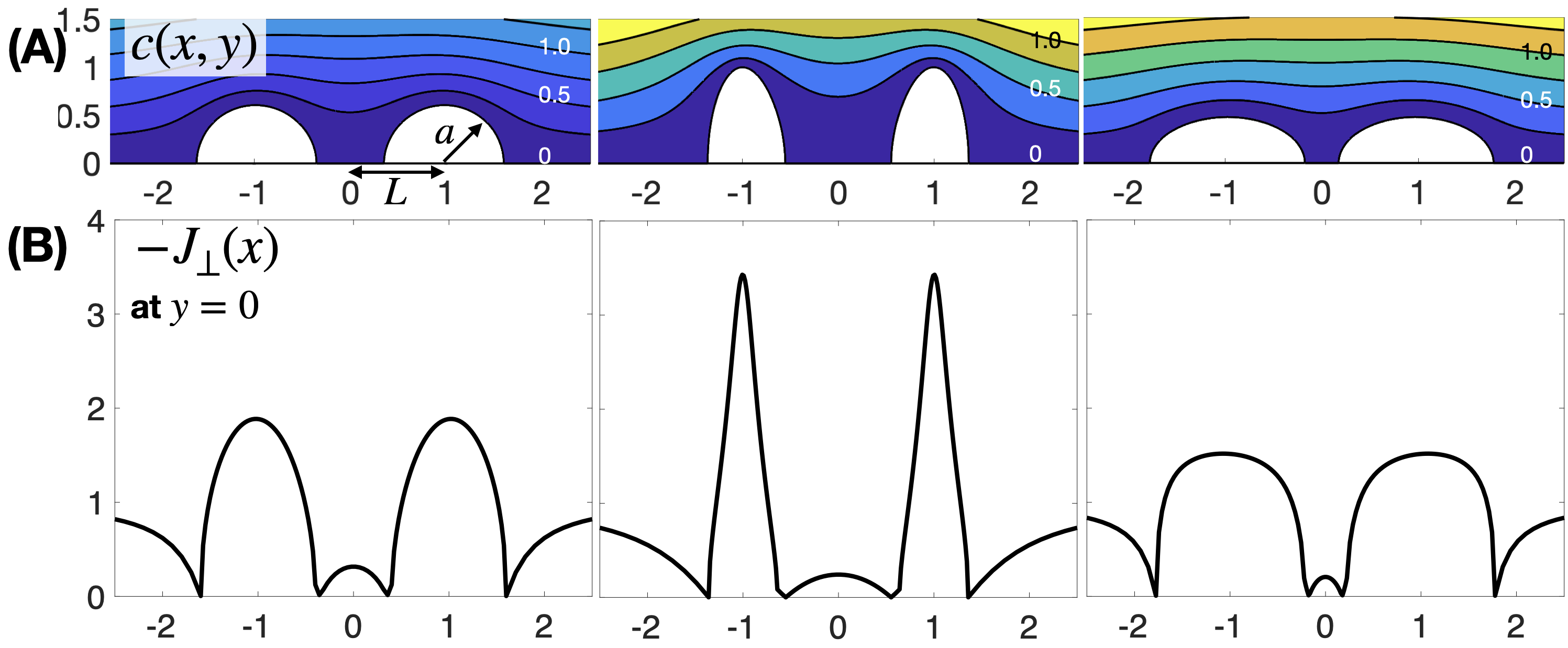}
    \caption{(A) Vapor concentration field and contours with two semicylinders ($a=0.6$ \& $L=1$), two prolate semi-ellipses ($a=0.68$ \& $b=0.47$ \& $L=1$), or two oblate semi-ellipses ($a=0.62$ \& $b=0.31$ \& $L=1$) from left to right. (B) Plots of diffusion flux onto three different surfaces. \red{ \sout{Here,}} The diffusion flux is normalized by the flux on a flat surface. High downward fluxes are found near the top of bumps. }
    \label{fig:flux_dipole}
\end{figure*}

\subsubsection{Two semicylinders}
The approximate solution can be obtained by placing two dipoles at the centers of two semicylinders. Then, the strength of each dipole can be determined by satisfying a boundary condition. This approximate solution can be further used for the case of more than two objects too. However, we will demonstrate the case with only two bumps in this paper. One caveat is that this method works well when the distance between two structures is larger than the size of the structures. 

The complex potential of a dipole can be written as $-a^2/z$ where $a$ is the radius of the semicylinder. When we place two semicylinders of radius $a$ at a distance $L$ from the center, the complex potential will become
\beq
W(z) =  z - A_1 \frac{a^2}{z-L} - A_2  \frac{a^2}{z+L} \,.
\eeq 
Since the bump shape is the same for both (i.e., symmetric across $x=0$), we can set two unknowns to one as $A_1 = A_2 \equiv {\cal A}_0$. This unknown, ${\cal A}_0$, can be determined from one boundary condition $\mathrm{Im}[W]|_{z = a+ L \mathrm{~ or~} a-L} = 0$. It does not matter whether you choose the right or left boundary condition. From the boundary condition, one can find ${\cal A}_0=-(a+2L)^2/((a+2L)^2+a^2)$. \red{Alternatively, we can write the potential by shifting the potential by $L$ as $W(z) = z + L - A_1 a^2/z - A_2 a^2/ (z+2L)$.  In the limit of $L \rightarrow \infty$, it converges to the potential with a single semicylinder. }

Then the complex potential becomes 
\beq
{W(z)} =  z +  \frac{(a+2L)^2 a^2}{(a+2L)^2+a^2} \, \frac{2z}{(z^2-L^2)}  \,.
\eeq 
One thing about this solution is that in the limit of $L\rightarrow 0$, this solution converges to the solution with a single semicylinder.  

The vapor flux is calculated from the derivative of the complex potential above.  
\beq
{J_\perp} = -\frac{dW}{dz} = -\left[ 1 - \left( \frac{(a+2L)^2 a^2}{(a+2L)^2+a^2} \right) \frac{2(z^2+L^2)}{(z^2-L^2)^2}  \right] \,.
\eeq 
Here, by replacing $z$ with a point along the surface, we can calculate the diffusion flux on the surface. It is a bit complicated to get analytical expression of diffusion flux on the bump. However, as you see the shapes and magnitudes of diffusion flux in Fig. \ref{fig:flux_dipole} and Fig. \ref{fig:bump}, they are in a similar shape and value. So, you can approximate the diffusion flux on each bump as the expression listed in Table \ref{tab:one}. 

\subsubsection{Two semi-ellipses}
Similar to the previous section, the approximate solution can be obtained by placing a certain form of the complex potential at the centers of two semi-ellipses. The complex potential to be used is the second term in Eq. (\ref{eq:cp_ellipse}). For two semi-elliptical bumps, the complex potential is composed of two of this term along with two unknowns. Similarly, we can set both unknowns are the same due to the structural symmetry. Each semi-ellipse has the major half-length as $a+b^2/a$ and the minor half-length as $a-b^2/a$. The distance from the center of each semi-ellipse to the origin $x=0$ is $L$.

Then the complex potential becomes 
\beqs
{W(z)} =  z - {\cal A}_1 \left( 1\pm \frac{a^2}{b^2} \right)   \left(z - \sqrt{\frac{(z-L)^2}{2^2} \pm b^2}  - \sqrt{\frac{(z+L)^2}{2^2} \pm b^2} \right) \,,
\eeqs 
where ${\cal A}_1 (a,b,L)$ is a constant given in Table \ref{tab:two}. Same as before, the plus sign is for a prolate shape and the minus sign is for an oblate shape. This constant is calculated to satisfy the boundary condition, $\mathrm{Im}(W)|_{z = (a\mp b^2/a) \,\pm L} = 0$.

From its derivative, the vapor flux is given as 
\beqs
{J_\perp} = -\frac{dW}{dz} = -\left[ 1 - {\cal A}_1 \left( 1\pm\frac{a^2}{b^2} \right)   \left(1 - \frac{\frac{(z-L)}{2^2}}{\sqrt{\frac{(z-L)^2}{2^2} \pm b^2}}  - \frac{\frac{(z+L)}{2^2}}{\sqrt{\frac{(z+L)^2}{2^2} \pm b^2}} \right) \right]  \,.
\eeqs


Figure \ref{fig:flux_dipole} shows the vapor concentration field and flux around two bumps. General features are similar to the case of a single bump: the concentration contours are pushed up by bump structures. As the structures are elongated vertically, the contours are packed and lifted up. Therefore, a higher flux is developed near the top of microsctructures. One interesting feature in the case of two bumps is that the flux value in between two bumps. It does not reach to the far-field limit ($J_\perp(x\rightarrow \pm \infty) = -1$), instead it creates a small flux region. This small flux region between the two bumps indicates that any condensed drops hardly grow when they are located in between bumps.

\subsubsection{Vapor from a stomatum with two semi-ellipses} 
\begin{figure*}[t]
    \centering
    \includegraphics[width=.8\textwidth]{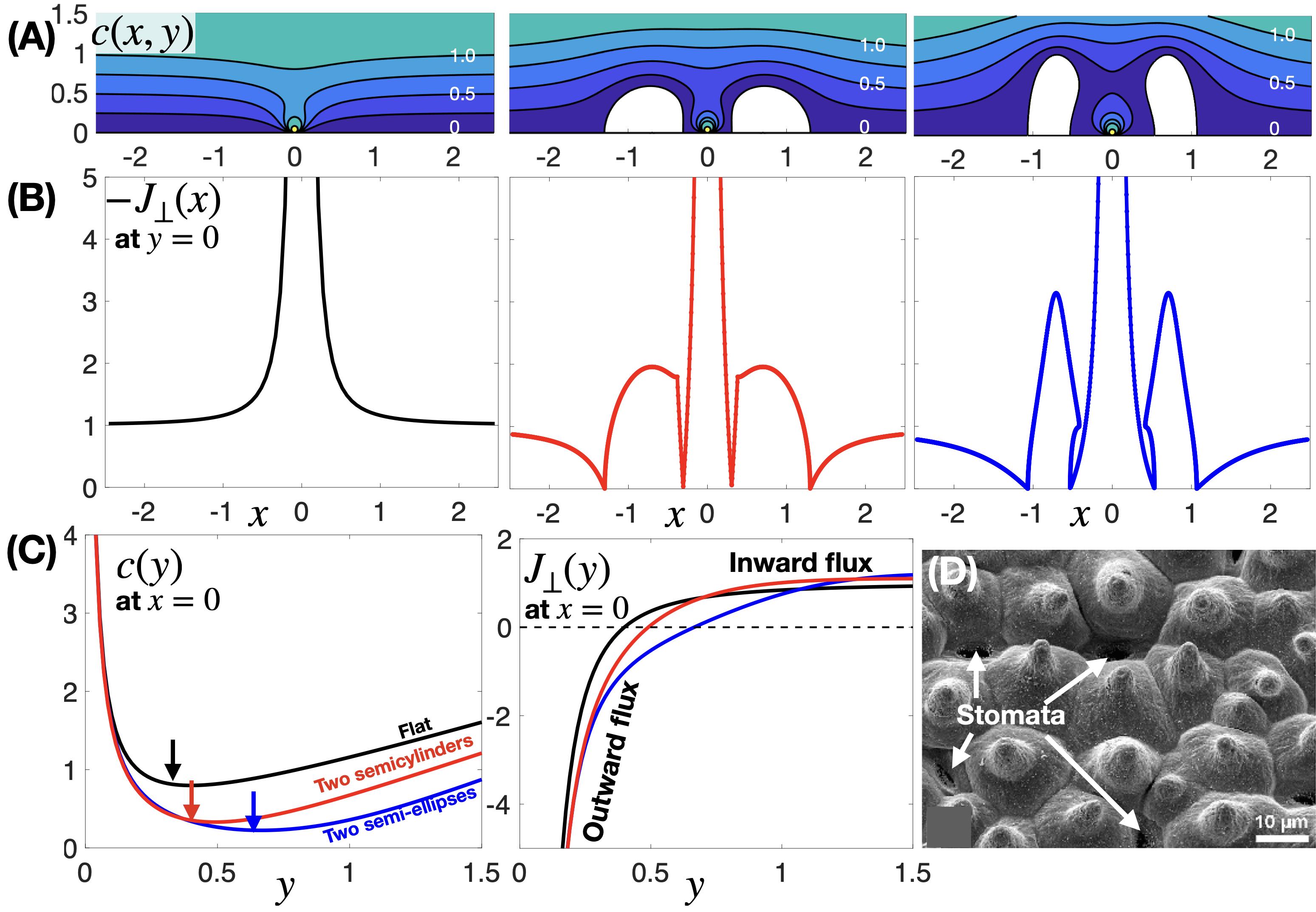}
    \caption{(A) Diffusion flux when the stomatum opens with a high vapor flux ($Q=1$) on flat, semicylindrical bumps ($a=0.6$ \& $L=0.7$), or  prolate bumps ($a=0.68$ \& $b=0.47$ \& $L=0.7$). (B) Diffusion flux on the leaf surface. A higher outward flux between two prolate bumps than one on a flat surface. (C) vapor concentration and flux along the centerline $x=0$. (D) The SEM image shows that the stomata of \textit{Nelumbo nucifera} are surrounded by epidermal bumps (\red{reproduced} from ref. \cite{Ensikat2011}). }
    \label{fig:flux_dipole_source}
\end{figure*}

We simulate the vapor escaping from a stomatum that is a small opening on a leaf surface. Typically, the inner space of a stomatum is fully saturated due to high water contents. Such high vapor concentration through the small opening will create a outward flux to diffuse the water molecules out. These stomata are surrounded by other epidermal bumps as shown in Fig. \ref{fig:flux_dipole_source}(D). To simplify this flux, we will place a small dipole pointing the horizontal direction in the middle of the two bumps. 

The dipole source to simulate the flux from a stomatum can be expressed as 
\beq
W_\mathrm{source} = - \frac{Q}{2 \pi z} \,.
\eeq 
Here, $Q$ is the strength of the dipole, i.e., the total flux. By adding this dipole source into a solution with two bumps, we can simulate a situation close to a stomatum in between two bumps as shown in Fig. \ref{fig:flux_dipole_source}.

For two semicylinders, the superposed complex potential will be 
\beqs 
W 
= z - {\cal A}_0 \frac{a^2}{z-L} - {\cal A}_0  \frac{a^2}{z+L} - \frac{Q}{2\pi z} 
\eeqs 
where ${\cal A}_0$ is a coefficient to be chosen to satisfy a boundary condition as 
\beq
\left. \frac{dW}{dz} \right|_{(z=a+L)} = 1+ {\cal A}_0 + {\cal A}_0  \frac{a^2}{(a+2L)^2} +  \frac{Q}{2\pi (a+L)^2}  = 0 \,. 
\eeq 
Hence, we get $ {\cal A}_0 = -({Q}/[2\pi (a+L)^2] +1)/(1+C)$ where $C \equiv a^2/(a+2L)^2$.

For two semi-ellipses with a stomatum source, the superposed complex potential will be 
\beqs
W(z) 
&=&  z - {\cal A}_1 \left( 1\pm \frac{a^2}{b^2} \right) \left(\frac{z-L}{2} - \sqrt{\frac{(z-L)^2}{2^2} \pm b^2}  \right)   \nonumber \\ 
&&- {\cal A}_1  \left( 1 \pm \frac{a^2}{b^2} \right) \left(\frac{z+L}{2} - \sqrt{\frac{(z+L)^2}{2^2} \pm b^2}  \right) - \frac{Q}{2\pi z} 
\eeqs 
where ${\cal A}_1$ is a coefficient to be chosen based on a boundary condition. 
That boundary condition is
\beqs
\left. \frac{dW}{dz} \right|_{z=a\mp b^2/a+L} &=&  1 - {\cal A}_1 \left( 1\pm \frac{a^2}{b^2} \right)  \left(1 - \frac{\frac{a \mp b^2/a}{2^2}}{\sqrt{\frac{(a \mp b^2/a)^2}{2^2} \pm b^2}}  - \frac{\frac{a \mp b^2/a+2L}{2^2}}{\sqrt{\frac{(a \mp b^2/a+2L)^2}{2^2} \pm b^2}} \right) \nonumber  \\
&&+ \frac{Q}{2\pi (a \mp b^2/a+L)^2}  = 0\,.
\eeqs 

Hence, we get 
\beq 
{\cal A}_1 = -\frac{1+ \frac{Q}{2\pi (a \mp b^2/a+L)^2} } { \left( 1 \pm \frac{a^2}{b^2} \right)  \left(1 - \frac{\frac{a \mp b^2/a}{2^2}}{\sqrt{\frac{(a \mp b^2/a)^2}{2^2} \pm b^2}}  - \frac{\frac{a \mp b^2/a+2L}{2^2}}{\sqrt{\frac{(a \mp b^2/a+2L)^2}{2^2} \pm b^2}} \right) } \,.
\eeq 
For the plus-minus or minus-plus signs, the upper sign is for a prolate shape and the lower sign is for an oblate shape.

Figure \ref{fig:flux_dipole_source}(A) demonstrates the case when a small opening in the middle ejects water vapor like open stomata on a flat surface (left image), on the surface with two semicylinders (middle image), and on the surface with two prolate semi-ellipses (right image). It is worth noting that the boundary of solid structures is deformed in the presence of the dipole. As mentioned earlier, this method of using superposed singularities works well with a large distance between dipoles. However, to demonstrate the case of an outward flux close to and in between two bumps, it is inevitable to get the bump contour distorted from ideal semicylinders or prolate shapes. Figure \ref{fig:flux_dipole_source}(B) shows a flux profile on a leaf surface. Similar to the case of a single bump, a higher flux is formed near the top of the flux. Also, the order of the magnitude and the shape of the flux is similar to the flux solution in a single bump. 

Figure \ref{fig:flux_dipole_source}(C) shows the vapor concentration and flux along the centerline above a stomatum ($x=0$). The vapor concentration profile shows a sharply decreasing trend starting from the stomatum and then a gradually increasing trend. As you know that the vapor flows from a high concentration region to a low concentration region. So, we expect the outward flux from the stomata and the inward flux from the atmosphere. The outward diffusion flux ($J_\perp > 0$ i.e., $\nabla c<0$) becomes zero in a distance away from the stomatum, in which the inward flux cancels out. Two fluxes in opposite directions balance out at some point in the air. As this zero-flux point gets away from the stomatum, more area with less vapor is available for the vapor to escape from the inner space of the leaf. As shown in the right panel of Fig. \ref{fig:flux_dipole_source}(C), the zero-flux point is moved away from the stomatum as the bumps have more prolate shape. Based on the image of contours in Fig. \ref{fig:flux_dipole_source}(B), prolate bumps will hinder the inward flux from the atmosphere and make room for the outward vapor flux from the stomatum. Therefore, the vapor from the stomata can easily diffuse out to the atmosphere in the presence of prolate bumps rather than the cases with cylindrical/oblate bumps or a flat surface.

\begin{figure*}
    \centering
    \includegraphics[width=.7\textwidth]{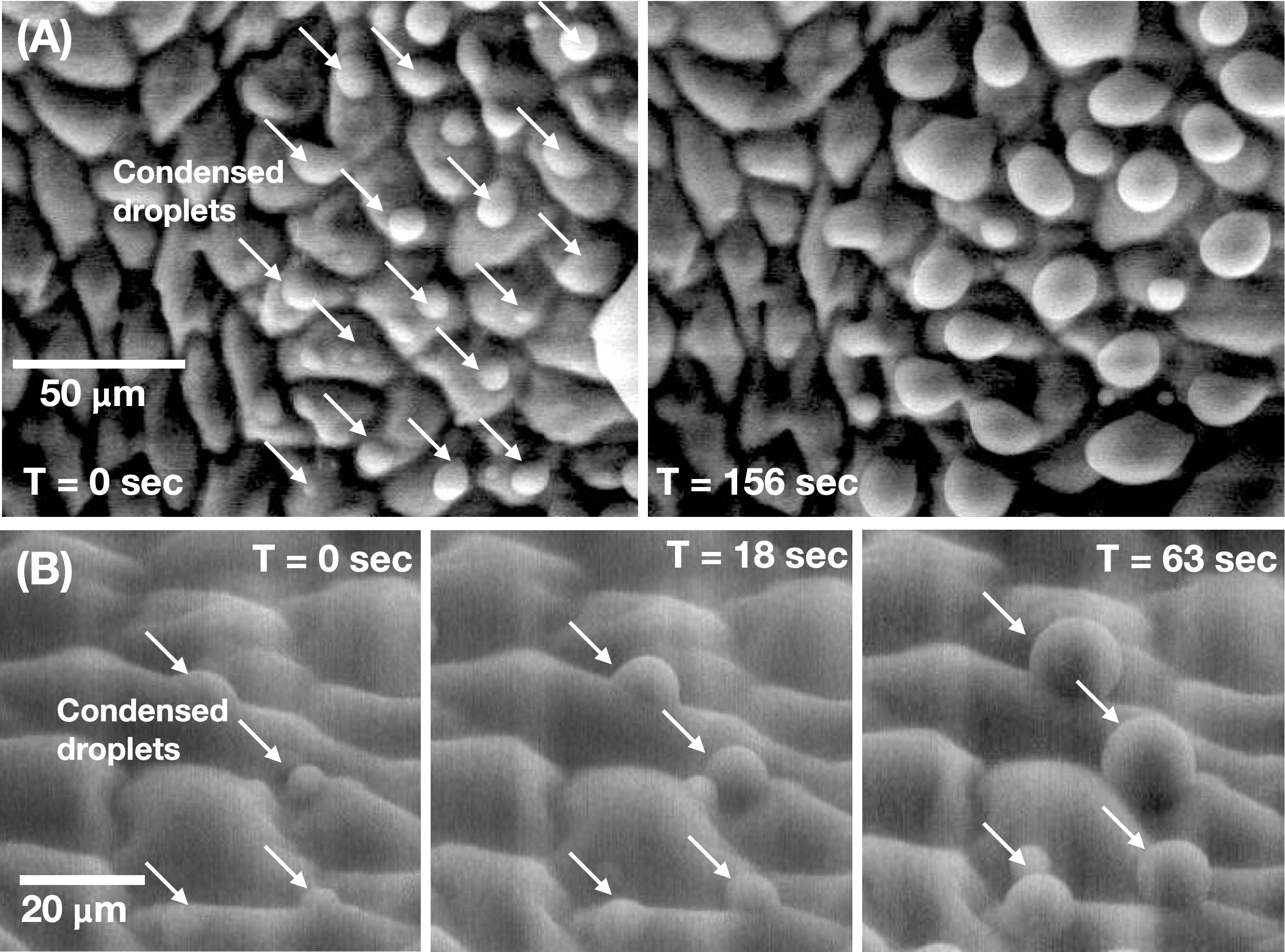}
    \caption{Drop condensation experiments on a Katsura leaf. (A) SEM images from the top view. (B) SEM images from the angled view. You can see growing droplets (pointed by arrows) only near the top of epidermal bumps.}
    \label{fig:my_label2}
\end{figure*}

\section{Conclusions} 

In this paper, we provide simple and analytical solutions of vapor concentration and diffusion flux on different surfaces: semicylinder, prolate semi-ellipse, oblate semi-ellipse, and hair-like structures. We find that a high inward diffusion flux (i.e., a high mass flux of vapor) is developed near the top of the microstructures, thereby droplets easily condense and grow. On the other hand, a low diffusion flux is formed near the stomata or the lower side at the leaf surface. Such a low vapor flux near the stomata could affect transpiration in two ways. First, the condensed droplets on the stomata will not grow due to a low mass flux of vapor, which lets the open stomata fully exchange gases. Second, the low vapor flux from the atmosphere will not hinder the release of high concentrated vapor from the substomatal space much. \red{Our results can be applicable to the concentration and vapor flux very close to the surface at the scale of microstructures. We did not consider a large-scale concentration gradient in this study. Additionally, if there is a slight air flow in real situations, we need to solve the advected diffusion equation, which is beyond our scope of the study. }


Currently, we do not have any quantitative measurements to compare experimental results with theoretical solutions \red{ \sout{here}}, but our preliminary results on the leaf surface show the likelihood of drop-wise condensation on the upper portion of the microstructures qualitatively as shown in Fig. \ref{fig:my_label2}. In the future, we will perform quantitative experiments to verify the solutions and develop surrogate analytical solutions in 3D bumps. Another interesting fact is that plant leaves are very dynamic due to raindrop impact \cite{Gart2015a,Bhosale2020,Kim2020} or wind \cite{Louf2018a,DeLangre2008}, which will further modify the vapor concentration and flux around. Additionally, even though we present our results from a plant transpiration perspective, many of these results can be useful to understand condensation or evaporation on nonuniform engineering surfaces. 


\section{Codes}
All matlab codes are freely available in DOI:10.17605/OSF.IO/PWVHE. Figure \ref{fig:bump}(B,C) are generated using Circle.m, Prolate.m, and Hair.m. \red{Figure \ref{fig:flux_max}(A,B) and Figure \ref{fig:flux_optimization}(B) are generated from Optimization.m.} Figure \ref{fig:flux_dipole}(A,B) are from Two\_bump.m. Figure \ref{fig:flux_dipole_source}(A,B,C) are from Two\_bump\_source.m.

\section{Acknowledgement}
The author thanks \red{Prof. Kyoo-Chul Park for providing data from his previous publication and} Dr. Hosung Kang for providing SEM images of drop condensation on a leaf. This work was partially supported by the National Science Foundation (Grant No. CBET-2028075 and CMMI-2042740).

\bibliography{NSF21}


\end{document}